\documentclass[11pt]{article}
\usepackage[paper=a4paper,dvips,top=2.5cm,left=2.5cm,right=2.5cm,bottom=2.6cm,footskip=2.2cm,voffset=0.0cm, marginparwidth=18mm]{geometry}
\usepackage{amsfonts,amsmath,amsthm,amssymb}
\numberwithin{equation}{section} 
\usepackage[svgnames]{xcolor}
\usepackage{fix-cm}    
\usepackage{sectsty}
\usepackage{fancyhdr}
\usepackage{lastpage}
\usepackage{amsmath}
\usepackage{graphicx}
\usepackage{url}
\usepackage{enumerate}
\usepackage{paralist}
\usepackage{multicol}
\usepackage{color}  
\usepackage{float}  
\usepackage{epsfig}
\usepackage{subfigure}
\usepackage{hyperref}   
\hypersetup{colorlinks=true,linkcolor=beamer@PRD, citecolor=beamer@PRD}
\usepackage{authblk}  
\usepackage{cite}  
\usepackage{todonotes}
\usepackage{ulem} % Enables sout command
\renewcommand\eqref[1]{\textcolor{beamer@PRD}{(}\ref{#1}\textcolor{beamer@PRD}{)}}
\definecolor{beamer@PRD}{RGB}{46,48,146}
\begin{document} 
\title { { Quantum deformation of cubic  string field theory }}
\author{\small{Arshid Shabir$^1$, Amaan A. Khan$^2$,  James Q. Quach$^{3}$, Salman Sajad Wani$^4$, Mir Faizal$^{5}$,   Seemin Rubab$^6$}
\\
\textit{\small $^{1, 5}$Canadian Quantum Research Center, 204-3002, 32 Ave Vernon, BC V1T 2L7, Canada}
\\
\textit{\small $^{2}$Irving K. Barber School of Arts and Sciences, University of British Columbia, Kelowna, British Columbia, V1V 1V7, Canada}
\\
\textit{\small $^3$CSIRO, Ian Wark Laboratory, Bayview Ave, Clayton, Victoria, 3168, Australia}\\
\textit{\small $^3$ The University of Adelaide, South Australia 5005, Australia}
\\
\textit{\small $^{4}$Center for Quantum Computing, Hamad Bin Khalifa University, Doha, Qatar}
\\
\textit{\small $^5$University of British Columbia - Okanagan, Kelowna, British Columbia V1V 1V7, Canada}
\\
\textit{\small $^6$Department of Physics,  National Institute of Technology,
Srinagar, Jammu and  Kashmir 190006, India}}
\date{}
\maketitle 
\begin{abstract}
In this paper, we will analyze a quantum deformation of cubic string field theory.   This will be done by first constructing a quantum deformation of string theory, in a covariant gauge, and then using the quantum deformed stringy theory to construct a  quantum deformation of string field theory. This quantum-deformed string field will then be used to contract a quantum-deformed version of cubic string field theory. We will explicitly demonstrate that the axioms of cubic string field theory hold even after quantum deformation.  Finally, we will analyze the effect of the quantum deformation of string field theory on the string vertices.
\end{abstract}

\section*{Keywords}
    Quantum Deformation, Cubic String Field Theory, String Vertices,  Witten Vertex.

\tableofcontents

%%%%%%%%%%%%%%%%%%%%%%%%%%%%%%%%%%%
\section{Introduction}
Several deformations of string theory have been studied, including p-adic string theory     \cite{q101, q7},   noncommutative deformation  \cite{q8,q1, q2},  $\kappa$-deformation \cite{q3}, $\Omega$-deformation  \cite{q4}, $T\bar T$-deformation \cite{q5a} etc. One of the most interesting deformations of string theory is the quantum (\textit{q})-deformation  \cite{kappa, first, first14}. Quantum Lie algebra is a quantum deformation of the usual Lie algebra, and the group corresponding to this quantum Lie algebra is called a quantum group. The  \textit{q}-deformed algebra reduces to the usual Lie algebra when the deformation parameter \textit{q} is set equal to unity.
Here,  the \textit{q}-deformed Lie algebra is the typical enveloping algebra deformed by means of the single parameter  \textit{q}.  In fact, such an algebra can be viewed as a form of Hopf algebra \cite{n0d, n1d, n2d}. It is clear that \textit{q}-deformed algebra provides us with a class of symmetries that are richer than the symmetries represented by Lie algebra. It is therefore conceivable that \textit{q}-deformed algebra can turn out to be appropriate for describing symmetries of physical systems, which are outside the realm of Lie algebras. This has motivated  the study of the mathematical structure of the quantum groups, which   has been
extensively explored in connection with several important aspects of physical phenomena.
Among them, is the deformation of conformal field theories, which plays 
an important role in understanding the structure of various important physical systems  \cite{l3}.       It is possible to study such deformation of a   conformal field theory using an algebraic approach based on 
the Yang–Baxter equation \cite{yangb}. The solution of the quantum Yang-Baxter equation for integrable models has been constructed using such a  \textit{q}-deformation of Lie algebra  \cite{Z2,Z4}.  The  Yang-Baxter equation has also been used to investigate string dualities \cite{dualties}. This involves both the T-duality and S-duality associated with string theories. The U-duality of M-theory has been studied using the Yang-Baxter equation \cite{mtheory}. This was done using the exceptional Drinfel’d algebra, which is a  Leibniz algebra. It is possible to analyze deformations of bosonic string theory using the  Yang-Baxter equation \cite{Baxter}. The  Yang-Baxter has been used to study the duality of twisted fields in the context of gauged double field theory \cite{Yang}. 
  
It is possible to investigate the properties of  \textit{q}-deformed algebra for various conformal field theories.
The quantum exchange algebra for the WZW model has been related to such a \textit{q}-deformed algebra, namely the quantum  $sl_q(2)$ algebra \cite{Z5}. The \textit{q}-deformed Ginsparg–Wilson algebra has been constructed by using \textit{q}-deformed pseudo-fuzzy Dirac and chirality operators  \cite{diracc}. A generalized \textit{q}-deformed symplectic algebra, namely  $sp_q(4)$ algebra has been used for multishell applications \cite{multi}. 
The  \textit{q}-deformation of the  Witt algebra has already been constructed \cite{Z6,Z7}. This \textit{q}-deformed Witt algebra has been generalized to  \textit{q}-deformed Virasoro algebra by including the central extension   \cite{Z8}. In the $q \to 1$ limit, this \textit{q}-deformed Virasoro algebra reduces to usual  Virasoro algebra.
It is known that the stress tensor in string theory obeys the commutation relations of two copies of the Virasoro algebra. Thus, the \textit{q}-deformation of the  Virasoro algebra naturally motivates the study of a \textit{q}-deformation of world-sheet string states. In fact, such 
 \textit{q}-deformation of first  quantized string theory has already been investigated \cite{kappa, first, first12, first14}. This was done from a  \textit{q}-deformation of the string oscillator algebra, such that this novel oscillator algebra coincided with the usual oscillator algebra in the $q \to 1$ limit. It has been demonstrated that such  \textit{q}-deformed algebra can be used to analyze the asymptotic behavior and
the spectrum of resonances of dual models for the strong interaction \cite{dual1, dual2}. 

The  \textit{q}-deformation of string theory has also been motivated by holographic duality.  A well-studied example of this  duality involves  the duality between  type IIB string theories on
the $ AdS_5 \times S^5$ background and 
$\mathcal{N} = 4$  Yang-Mills theory in the large $N$ limit \cite{kl, lk}. 
An interesting aspect of this duality is related to the integrability of the world-sheet theory. It is possible to obtain magnon-like solutions for strings  \cite{inte0, inte01}. Using the holographic duality and integrability, it is possible to express the  S-matrix for such magnon-like solutions \cite{inte1}. The   S-matrix of this theory has a Yangian symmetry associated with centrally extended Lie superalgebra, which can be deformed using quantum groups.  Here, the   \textit{q}-deformation of the S-matrix is done by suitable replacing the rational R-matrix describing the magnons    \cite{inte2}. This  \textit{q}-deformation of magnons is the 
quantum analog of the classical result involving the continuous deformation of the symplectic structure of integrable systems.  This deformed theory   depends both on  the   coupling of the original  theory $g$  (which is the ’t Hooft coupling of the dual theory)  and the   \textit{q}
parameter. The original magnon theory is obtained in the  $\textit{q} \to 1$ limit. However, it is possible to take $g \to \infty$  and keep \textit{q}   fixed \cite{inte4}. This leads to an S-matrix for a  theory, which is classically equivalent to the string world-sheet theory \cite{inte5, inte6}.  The S-matrix for generic $g$ and \textit{q} has been constructed, and it has been observed that it is important to consider the dressing phase in the stringy limit  \cite{smatrix, smatrix1}. Here, it is important to consider both the original magnon theory and the mirror theory, which is obtained from a double Wick
rotation of the world-sheet string theory \cite{intea1, intea2}. 
The   \textit{q}-deformation of string theory on  $AdS_5 \times S^5$ has also been holographically investigated \cite{xz12, XZ}.   

Various other important holographic results have been obtained using the  \textit{q}-deformation of the string theory on  $AdS_5 \times S^5$ \cite{B1,C1}. In fact,  string theory on a \textit{q}-deformed   $AdS_5 \times S^5$  has been used to analyze the corresponding deformed field equations, Lax connection, and $\kappa$-symmetry transformations \cite{kappa}.   The   \textit{q}-deformation of string theory on $AdS_3 \times S^2 \times T^4$ and its mirror duality has been investigated \cite{T4}. Here,  the exact \textit{q}-deformed S-matrix was obtained for light-cone string theory on this background.  The behavior of this \textit{q}-deformed S-matrix under mirror transformation was investigated using a double Wick rotation of the world sheet. 
The holographic  Wilson loops in \textit{q}-deformed $AdS_5\times S^5$ geometry have also been studied \cite{wilson}. Here, a  minimal surface in this geometry is dual to a cusped Wilson loop of the dual field theory. The area of the minimal surface contains logarithmic squared divergence and logarithmic divergence. The string theory on a   \textit{q}-deformed  $AdS_5 \times S^5$  has been studied using the classical integrable structure of anisotropic Landau-Lifshitz sigma models \cite{sigma}. Thus, \textit{q}-deformation of string theory is also been motivated by holography. 

 Motivated by the \textit{q}-deformation of first quantized string theory \cite{first}, it is interesting to analyze the \textit{q}-deformation of second quantized string theory.  In analogy with field theory, it is possible to obtain a  Feynman diagram-like expansion for string scattering amplitudes in string field theory by using vertices for joining strings and constructing suitable string propagators.   This is done by constructing a string field theory, which satisfies a set of axioms \cite{Erler:2019loq}.
 As string field theory, is based on a  field theory formalism,  it can be used to investigate off-shell string amplitude \cite{shell1, shell2}. 
 Another advantage of using string field theory is that it is a background-independent formalism \cite{back1, back2}. The divergences that occur due to 
  a Euclidean world-sheet description for the boundary of moduli space can be resolved in string field theory \cite{sft01, sft02}.  The string field theory also resolves the problems relating to the unitarity \cite{prob6} and crossing symmetry \cite{prob7} of scattering amplitudes in string theory.  Thus, the string field theory resolves several problems that occur with the first quantized string theory. 
  
 It has been demonstrated that these axioms are satisfied by Witten's open string field theory, which is a string theory with cubic interaction   \cite{witten}. 
 We would like to point out that the deformation of Witten's open string field theory is expected to produce important changes in the behavior of this system. Witten's string field theory has been deformed to map    the non-planar world-sheet diagrams of the perturbative string theory to their equivalent planar diagrams in the light-cone
gauge (with some fixed parameters) \cite{lightcone}. It was argued that the scattering amplitudes with an arbitrary number of open strings could be investigated using such a deformation. It may be noted that various other deformations of string field theory,   such as $\mu$-deformation \cite{mu},  sine-square deformation  \cite{sine}, marginal boundary deformation of background Wilson lines \cite{margin}, mass-term deformation \cite{mass},  marginal deformation from the tachyon vacuum \cite{defo1},   non-planar deformation  \cite{defo2}  have been studied. In all these deformations, the spectrum of string field theory changes, but the quantum algebra of the creation and annihilation operators is not deformed and is represented by the standard quantum algebra. Thus, \textit{q}-deformation of string field theory has not been constructed even though the  \textit{q}-deformation of first quantize string theory has already been constructed  \cite{kappa, first,  first14}. This deformation would be fundamentally different from other deformations of string field theory, as it would involve a fundamentally different commutation algebra between the creation and annihilation operators. 
 It may be noted that   \textit{q}-deformation occurs in many-system interactions \cite{inte}, and as string field theory involves many fields, it is interesting to analyze its  \textit{q}-deformation. So,  we will investigate the \textit{q}-deformation of Witten's string field theory. This will be done by first constructing the \textit{q}-deformation of all the standard ingredients of Witten's open string field theory, and then using those \textit{q}-deformed ingredients to 
construct a \textit{q}-deformation of Witten's open string field theory. 

\section{Witten's open string field theory}
In this section, we review Witten's open cubic string field theory \cite{witten}. We formally construct this string field theory using the language of functional integrals. We describe the matter sector of the string field in terms of the delta function interaction.
Witten's open string field theory is based on  an  action of Chern-Simons form \cite{witten}
\begin{equation}\label{action}
    S= -\frac{1}{2}\int \Psi \star Q\Psi - \frac{g}{3}\int\Psi \star\Psi \star\Psi,
\end{equation}
where $g$ is the open string coupling, $\Psi$ is the string field taking the values in graded algebra $\mathcal{A}$. And this algebra $\mathcal{A}$ is equipped with non-commutative star product  
\begin{equation}
    \star: \mathcal{A} \otimes  \mathcal{A} \rightarrow \mathcal{A},
\end{equation}
under which the degree $G$ is additive: $(G_{\Psi \star \Phi} = G_{\Psi} + G_{\Phi})$. Also a BRST operator 
\begin{equation}
    Q : \mathcal{A} \rightarrow \mathcal{A},
\end{equation}
of degree one: $(G_{Q\Psi}= 1+ G_{\Psi})$.
This $\star$ is used to define a product of string fields in string field theory.  It encodes all the interactions by specifying how two incoming strings are glued into the resulting one. Here, $Q$ is the usual open-string BRST operator.
\begin{equation}
    Q= \sum_{n} c_n L^X_{-n}  +  \frac{1}{2} \sum_{n,m}\ :c_{m} c_{n} b_{m+n}: \ -ac_0,
    \end{equation}
where a is the normal ordering constant for $L_0$  and $L_m$ are the Virasoro generators given by \begin{equation}\label{VG}
   L_m=\frac{1}{2}\sum_{n=1} ^\infty \alpha^{\mu}_{m-n} \alpha^{\mu}  _n.
\end{equation}
The string oscillator operators $\alpha^{\mu}_{-n}$ is the creation and  $\alpha^\mu_n$ is the annihilation operators, which obeys the commutation relation
\begin{equation}\label{alpha}
    [\alpha^{\mu}_m, \alpha^{\nu}_{-m}] =m  \eta^{\mu \nu},
\end{equation}
where $\eta^{\mu\nu}$ is the target space metric and `m' can have any non-zero value.
The ghost creation and annihilation operators satisfy an anti-commutation relation.
\begin{equation}\label{ghosts}
    \{c_n,b_m\}=\delta_{n+m,0}\ \ ,\ \ \{c_n, c_m\}=\{b_n,b_m\}=0.
\end{equation}
The  ghost and anti-ghost fields on the open string satisfy periodic boundary conditions as
\begin{equation}
    c^{\pm}(\sigma +2\pi)=c^{\pm}(\sigma) , \ \ b_{\pm}(\sigma +2\pi)=b_{\pm}(\sigma).
\end{equation}
Under mode decomposition, these fields can be written as
\begin{eqnarray}
    c^{\pm}(\sigma)=\sum_n c_n e^{\mp in\sigma}\nonumber \\
     b_{\pm}(\sigma)=\sum_n b_n e^{\mp in\sigma}.
\end{eqnarray}
Here,  $Q, \star, \int$   are the main defining objects of string field theory, and are assumed to satisfy the following axioms \cite{witten}:
\begin{itemize}
\item[] {(a)}   Nilpotency:\ $Q^2\Psi =0, \  \ \forall \Psi \in \mathcal{A}$.

\item[] {(b)} Integration:  $\int Q\Psi=0,  \  \ \forall \Psi \in \mathcal{A}$ .

\item[] {(c)}  Derivation: $Q(\Psi \star \Phi)=(Q\Psi)\star \Phi +(-1)^{G_{\Psi}}\  \Psi\star (Q\Phi), \  \  \forall \{\Psi, \Phi\} \in \mathcal{A}$.
 
\item[] {(d)} Cyclicity: $\int \Psi \star \Phi= (-1)^{G_{\Psi}G_{\Phi}} \int \Phi \star \Psi, \  \ \forall  \{\Psi,\Phi  \}\in\mathcal{A}   $.

\item[] {(e)} Associativity: $(\Psi \star \Phi) \star \Xi= \Psi \star (\Phi \star \Xi)$
\end{itemize}

These axioms are satisfied when  $\mathcal{A}$ is taken to be the space of string fields 
\begin{equation}
    \mathcal{A}=\{\Psi[X(\sigma);c(\sigma),b(\sigma)]\}.
\end{equation}
The string field is functional on matter, ghost, and anti-ghost field describing an open string ($0\leq \sigma \leq \pi$). Using such a string field, we can show that the  action \eqref{action} satisfies these axioms, and using these axioms, we can also demonstrate that it is invariant under the gauge transformations  
\begin{equation}
    \delta \Psi = Q\Lambda +\Psi \star \Lambda -\Lambda \star \Psi,
\end{equation}
for any gauge parameter $\Lambda \in \mathcal{A}$.

In conventional quantum field theory, elementary particles correspond to a local field (and vice versa). As there is an infinite number of elementary modes (or particles) in strings, it is natural to associate an infinite tower of usual fields with them. It turns out that all the spacetime  fields appear as
different coefficients in expansion of a string field $\Psi$ = $\Psi[X(\sigma), c(\sigma), b(\sigma)]$. 
This expansion in the Fock representation for a string field contains a family of spacetime fields, such as the scalar field $\phi(p)$, the vector field $A_\mu(p)$, and  higher spin fields, (along with coefficient fields for $b_n, c_n$ such as $\chi(p)$), and    can be written as \cite{witten, hh}
\begin{equation}
    \Psi =\int d^{26}p[\phi(p)|0;p\rangle + A_\mu(p) \alpha{_{-1}} |0;p\rangle+ \chi(p)b_{-1}c_0|0;p\rangle + \dots].
\end{equation}
Here, the vacuum state $|0;p\rangle$  is the total vacuum state of the system given by  the tensor product of quantities as:
\begin{equation*}
    |0;p\rangle= |0;p\rangle_{matter} \  \otimes \ |0;p\rangle_{ghost}.
\end{equation*}
where the vacuum state $|0;p\rangle_{matter}$ is of matter sector and $|0;p\rangle_{ghost}$ of ghost sector.
The $\star$ product is defined by gluing the right half of one string to the left half of the other using a delta function interaction. This interaction factorizes into separate matter and ghost sectors. Also integral over string field gets factorized into matter and ghost parts, and in the matter sector we can write 
\begin{equation}\label{lbl1}
    \int \Psi =\int \prod_{0\leq \sigma \leq \pi} dx(\sigma) \prod_{0\leq \sigma \leq \frac{\pi}{2}} \delta[x(\sigma)-x(\pi-\sigma)]\Psi[x(\sigma)].
\end{equation}
It is possible to define the ghost sector in a similar way. However, it should be noted that the ghost sector has an anomaly due to the curvature of the Riemann surface describing the three-string vertex. It can be described either in terms of fermionic ghost fields $c(\sigma), b(\sigma)$ or through bosonization in terms of the single bosonic scalar field $\phi(\sigma)$, from the functional point of view it is easiest to describe the ghost sector in the bosonized language. So, the integration of a string field in the ghost sector is given by   \eqref{lbl1} with an insertion of $\exp(-3i\phi(\pi/2))/2)$ inside the integral.
Now using a  bilinear inner product $\langle .,.\rangle$ on the state space of the conformal field theory describing the string, we can 
  algebraically write the open string field theory  action as 
\begin{equation}
    S= -\frac{1}{g^2}\Big[\frac{1}{2}\langle \Psi_1,{Q}\Psi_2\rangle + \frac{1}{3}\langle \Psi_1,\Psi_2 \star \Psi_3 \rangle\Big].
\end{equation}
 In this section, we reviewed the construction of Witten's open-string field theory. In the next section, we will analyze its \textit{q}-deformed version. 

\section{Quantum deformation}

 As the \textit{q}-deformation of string theory has been studied, we will construct a \textit{q}-deformation of Witten's cubic string field theory. 
Now we will investigate the \textit{q}-deformation of the  $\alpha$-modes and ghost/antighost modes by deforming Eqs.    \eqref{alpha} and  \eqref{ghosts}.
The \textit{q}-deformation of the string oscillator modes can be obtained using the  Biedenharn formalism   \cite{Biedenharn}. So, the \textit{q}-deformed version of Eq.  \eqref{alpha} can be expressed as 
\begin{equation}
   [\alpha^{\mu}_{m_q}, \alpha^{\nu}_{-m_q}]\equiv \alpha^{\mu}_{m}\alpha^{\mu}_{-m} -q \alpha^{\mu}_{-m}\alpha^{\mu}_{m} =[m]_q q^{-N_m}  \eta^{\mu \nu} ~,
\end{equation}
where $N_m$ is the \textit{q}-number operator corresponding to the mode $m$, and is defined by 
\begin{eqnarray}\label{qnuop}
   [N_{m_q}, \alpha^{\mu}_{-m_q}] = \prod_{m>0} [m]_q\alpha^{\mu}_{-m} &&
   [N_{m_q}, \alpha^{\mu}_{m_q}] = - \prod_{m>0} [m]_q\alpha^{\mu}_{-m}~.
\end{eqnarray}
And \textit{q}-number $[m]_q$ is defined to be 
\begin{equation}
    [m]_q= \frac{1-q^m}{1-q}= 1+q+\dots +q^{m-1}~.
\end{equation}
Similarly, the   \textit{q}-deformed anti-commutation relations of ghost modes are given by 
\begin{equation}\label{anti-ghost}
  \{{c_{m_q}},{b_{n_q}}\} =q^{-N_{m,n}}\delta_{m+n,0} \ , \ 
\{{c_{m_q}},{b_{n_q}}\}=\{{b_{n_q}},{b_{n_q}}\}=0 ~.
\end{equation}
Under the limit $q \rightarrow 1$ both \textit{q}-deformed string oscillator commutation relations as well as \textit{q}-deformed ghost modes anti-commutation relations reduce to their un-deformed versions given in Eqs.  \eqref{alpha} and \eqref{ghosts}.

The Virasoro algebra is an infinite-dimensional Lie algebra, which is important to analyze the world-sheet conformal field theory \cite{vira string}. 
The  \textit{q}-deformation of Virasoro algebra has already been investigated  \cite{V1,V2,V3}. A
deformation of the Virasoro algebra without the central charge (Witt algebra) has been obtained using the Curtright-Zachos procedure \cite{V1}. The deformation of Witt algebra with central charge extension has been done using the Jacobi identity \cite{V4}.
Now, we will analyze certain properties of the \textit{q}-deformed Virasoro algebra. The Virasoro generators given in Eq. \eqref{VG} have been  \textit{q}-deformed   \cite{NGT} as 
\begin{equation}\label{lnq}
    {L_{n_q}}  = \frac{1}{2} \sum_{n=1}^{\infty} \ :\alpha^{\mu}_{n-m} \alpha_{n} ^{\mu}:_q ~,
\end{equation}
where the \textit{q}-deformed normal ordering $ : \ :_q $ is defined as
\begin{eqnarray} \label{noq}
    :\alpha^{\mu}_{m_q} \alpha^{\nu}_{-m}:_q &\equiv& \alpha^{\nu}_{-m} \alpha^{\mu}_{m} +(q^{N_m}-1)\eta^{\mu \nu} \alpha^{\mu}_{-m} \alpha^{\nu}_{m} ~, \nonumber \\
    :\alpha^{\mu}_{-m} \alpha^{\nu}_{m}:_q &=& \alpha^{\mu}_{-m} \alpha^{\nu}_{m}~.
\end{eqnarray}

Given the \textit{q}-deformed oscillator modes and their commutation relation, let's compute the commutator of the \textit{q}-deformed Virasoro generators $( L_{n_q} )$:

The quantum deformation of the Virasoro algebra is obtained by generalizing the algebra $SU_q(1,1)$ which is a deformation of the universal enveloping algebra of $SU(1,1)$ since $SU(1,1)$ is a subalgebra of the Virasoro algebra. However, only two generalizations of it to q-Virasoro algebra have been achieved \cite{V1,ar2}. As central Virasoro algebra is a very important symmetry in conformal field theory as well as string theory, however, q-Virasoro algebra is important to examine the central extension of CZ-algebra.   
The  \textit{q}-deformed Virasoro algebra upon introducing an operator S has the structure     \cite{VA}
 \begin{equation}
     [{L_{n_q}}, {L_{m_q}}]=(n-m)L_{(n+m)_q} +C_q(n,m),
 \end{equation}
 and not in 
where $C_q(n,m)$ is the \textit{q}-deformed central extension which corresponds to a shift of a vector space and is given as;
\begin{equation}
    C_q(n,m)=q^{-4S}\frac{[n]}{[2n]} c \frac{[n-1][n][n+1]}{[3]_q!}\delta_{n+m,0}~
\end{equation}
with \textit{q}-deformed central charge is $c=(q^m +q^{-n})-C (n,-m)$ and $[n]_q! =[n]_q[n-1]_q \dots [1]$. The classical  limit of $C_q(n,m)$ is the same as the usual central charge of the Virasoro algebra;

\begin{equation}
    \lim_{q\rightarrow 1}C_q(n,m)=\frac{c}{12}(n^3-n) \delta_{n+m,0}~.
\end{equation}
The correspondence between the \textit{q}-deformed Virasoro algebra and the \textit{q}-deformed oscillator modes shows that the algebraic relations satisfied by the \textit{q}-deformed Virasoro generators are consistent with those of the \textit{q}-deformed oscillator modes.
The \textit{q}-deformed  BRST operator  can be written as 
\begin{equation}
    Q_q=\sum_{m>0}  L^X_{-m_{q}} c_{{m}_q}  -\frac{1}{2} \sum_{m>0}\sum_{n=1}^{\infty}(m-n):c_{-m} c_{-n} b_{m+n}:_q  \; -ac_0  ~.
\end{equation}
This expression can be written as:
\begin{equation}\label{abc1}
 Q_q=\sum_{m>0} \; :( L^X_{-m}   +\frac{1}{2}L^{gh}_{-m}  -a\delta_{m,0})c_m :_q
\end{equation}
The \textit{q}-deformed ghost number becomes;
\begin{equation}\label{ghost}
    U_q= \sum_{m>0} \; : c_{-m} b_m :_q
\end{equation}

 The  normal ordering of \textit{q}-deformed ghost operators $ : c_m c_n b_{-m-n}  :_q $ is defined in a similar fashion as of Eq. \eqref{noq}, 
\begin{eqnarray}\label{BCM}
   :c_{-m} c_{-n} b_{m+n}:_q  &\equiv& b_{m+n}c_{-m} c_{-n} -(1-{q^{-N_{m}}}) b_{m+n}c_{-m} c_{-n}~,  \nonumber \\
   :b_{m+n}c_{-m} c_{-n} :_q &=& b_{m+n}c_{-m} c_{-n} ~. 
\end{eqnarray}
Similarly, we can define the normal ordering of \textit{q}-deformed ghost number for Eq.\eqref{ghost} as:
\begin{eqnarray}\label{BCM}
   :c_{-m}  b_{m}:_q  &\equiv& b_{m}c_{-m}  -(1-{q^{-N_{m}}}) b_{m}c_{-m} ~,  \nonumber \\
   :b_{m}c_{-m} :_q &=& b_{m}c_{-m}  ~. 
\end{eqnarray}
Now, let's compute the $Q^2_q$;
\begin{equation}\label{qq2}
    Q^2_q = \frac{1}{2}\{ Q_q, Q_q\}
    = \sum_{m>0}\sum_{n=1}^{\infty} \left([ L_{m_q}, L_{n_q}] -(m-n)L_{(m+n)_q}\right) \; : c_{-m} c_{-n}:_q
\end{equation}
where $ L_{m_q} = L^X_{m_{q}} + L^{gh}_{m_{q}} - a\delta_{(m,0)}$ is a total \textit{q}-deformed Virasoro operator. Since the total central charge vanishes, $Q^2_q=0$ for critical dimensions $d=26$ and $a=1$. The inverse is also true: from $Q^2_q=0$ follows that the central charge of Virasoro algebra vanishes. First using Eqn, \eqref{anti-ghost}, we see that;
With the given expression for \(Q_q\) Eq. \eqref{abc1}:
and the \textit{q}-deformed commutation relations:
\begin{equation}\label{qwe2}
    \{{c_{m_q}}, {b_{n_q}}\} = q^{-N_{m,n}}\delta_{m+n,0}
\end{equation} 

We can attempt to calculate the desired Poisson bracket \(\{ Q_q, b_{m_q}\}\).
Let's proceed with the calculation:

\begin{equation}
    \{ Q_q, b_{m_q}\} = \sum_{n>0} \; :( L^X_{-n} + \frac{1}{2}L^{gh}_{-n} - a\delta_{n,0})\{c_n, b_m\} :_q 
\end{equation} 

Now, using the \textit{q}-deformed commutation relation Eq. \eqref{qwe2}: 
we get:

\begin{equation}
    \{ Q_q, b_{m_q}\} = \sum_{n>0} \; :( L^X_{-n} + \frac{1}{2}L^{gh}_{-n} - a\delta_{n,0})q^{-N_{n,m}}\delta_{n+m,0} :_q 
\end{equation}

Here, we observe that the sum is over \(n > 0\), and the term \(q^{-N_{n,m}}\) in the sum implies that only the term with \(n = m\) survives. Therefore, the expression simplifies to:

\begin{equation}
    \{ Q_q, b_{m_q}\} = : ( L^X_{-m} + \frac{1}{2}L^{gh}_{-m} - a\delta_{m,0}) :_q
\end{equation}

This result for the \textit{q}-deformed Poisson bracket \(\{ Q_q, b_{m_q}\}\) is basically equal to $  L_{m_q}$. So, we can write:
\begin{equation}
     L_{m_q} = \{ Q_q, b_{m_q}\}
\end{equation}
Next
\begin{align}
    [L_{m_q},Q_q]&= [\{Q_q, b_{m_q}, Q_q\}]=(Q_q b_{m_q} + b_{m_q} Q_q)Q_q -Q_q(Q_q b_{m_q}+ b_{m_q} Q_q)\nonumber\\
    &=b_{m_q} Q^2_q -Q^2_q b_{m_q}=[b_{m_q}, Q^2_q]
\end{align}
As $Q^2_q$ goes to zero, $[L_{m_q}, Q_q]$ also goes to zero. Therefore
\begin{align}\label{qq1}
[L_{m_q}, L_{n_q}] & = [L_{m_q}, \{Q_q, b_{n_q}\}] \nonumber\\
& = L_{m_q} Q_q b_{n_q} + L_{m_q} b_{n_q} Q_q - Q_q b_{n_q} L_{m_q} - b_{n_q} Q_q L_{m_q}\nonumber \\
& \quad + L_{m_q} Q_q b_{n_q} - Q_q L_{m_q} b_{n_q} + L_{m_q} b_{n_q} Q_q - b_{n_q} L_{m_q} Q_q \nonumber\\
& \quad + Q_q L_{m_q} b_{n_q} - Q_q b_{n_q} L_{m_q} + b_{n_q} L_{m_q} Q_q - b_{n_q} Q_q L_{m_q} \nonumber\\
& = [L_{m_q}, Q_q] b_{n_q} + [L_{m_q}, b_{n_q}] Q_q + Q_q [L_{m_q}, b_{n_q}] + b_{n_q} [L_{m_q}, Q_q] \nonumber\\
& = \{[L_{m_q}, Q_q], b_{n_q}\} + \{Q_q, [L_{m_q}, b_{n_q}]\}\nonumber \\
& = \{Q_q, [L_{m_q}, b_{n_q}]\} \nonumber\\
& = \{Q_q, (m-n){b_{(m+n)}}_q\} \nonumber\\
& = (m-n) \{Q_q, {b_{(m+n)}}_q\} \nonumber\\
& = (m-n) {L_{(m+n)}}_q.
\end{align}

Now, upon substitution of the result of Eq.\eqref{qq1} into Eq.\eqref{qq2}, we have:
\begin{align}\label{pop4}
    Q^2_q &=  \sum_{m>0}\sum_{n=1}^{\infty} \left([ L_{m_q}, L_{n_q}] -(m-n)L_{(m+n)_q}\right) \; : c_{-m} c_{-n}:_q\nonumber \\
    &= \sum_{m>0}\sum_{n=1}^{\infty} \left((m-n) L_{(m+n)_q} -(m-n)L_{(m+n)_q}\right) \; : c_{-m} c_{-n}:_q\nonumber \\
    &=0
\end{align}
The field theoretical formulation is based on the fact that Mandelstam-Neumann vertex functions are represented by $\delta$-function interaction \cite{i1,i2,i3}. This can be used to construct a \textit{q}-deformed second-quantized theory.  
The integral over string field factorises into matter and ghost parts, and in the matter sector, it is given by
\begin{equation}\label{;;}
 \int \Psi=\int \prod_{0\leq \sigma\leq\pi}dX{(\sigma)} \prod_{0\leq \sigma\leq \frac{\pi}{2}}\delta [X{(\sigma)} - X{(\pi - \sigma)}]  \Psi[X{(\sigma)}]~.
\end{equation}
Even though the infinite product of delta functions over a finite interval might have merely formal significance, this expression is given a specific meaning in terms of Fourier modes on the string.
However it's \textit{q}-deformed version can be expressed as;
\begin{equation}\label{01}
 \int \Psi_q=\int \prod_{0\leq \sigma\leq\pi}d_qX{(\sigma)} \prod_{0\leq \sigma\leq \frac{\pi}{2}}\delta_q [X{(\sigma)} - X{(\pi - \sigma)}]  \Psi_q[X{(\sigma)}]~,
\end{equation}
where \textit{q}-derivative $d_qX(\sigma)$ also known as the Jackson derivative, is defined as 
\begin{equation}
    d_q X(\sigma) = \frac{X(q^{N_m} \sigma)-X(q^{-N_m}\sigma)}{(q^{N_m}-q^{-N_m})X}
\end{equation}
Here, \(q\) is the \textit{q}-deformation parameter. This derivative captures the q-modifications to the rate of change of \(X(\sigma)\) as \(\sigma\) varies. When \(q\) approaches 1, the Jackson derivative reduces to the ordinary derivative.

This reduces to the plane derivative when $q\rightarrow 1$. The modified Dirac delta function called a \textit{q}-delta function can be written as
\begin{eqnarray}
    \delta_q [X{(\sigma)} - X{(\pi -   \sigma)}]=(1+q^{N_m}) \delta[X{(\sigma)} - X{(\pi -   \sigma)}] \nonumber \\ +2q^{N_m}X(\sigma)\frac{\partial}{\partial x}\delta [X{(\sigma)} - X{(\pi -   \sigma)}] \nonumber \\
    +\frac{q^{N_m}}{2}(X(\sigma))^2\frac{\partial^2}{\partial x^2} \delta[X{(\sigma)} - X{(\pi -   \sigma)}]~.
\end{eqnarray}
In the
limit  $q \rightarrow 0$, the \textit{q}-delta function reduces to the ordinary delta function. Here, the 
 \textit{q}-deformed string field is given in Eq. \eqref{psi}. 
 Similar definitions apply to the \textit{q}-deformed ghost sector of the theory, but it has an anomaly because of the curvature of the Riemann surface characterizing the three-string vertex. Either q-fermionic fields $c_q(\sigma), b_q(\sigma)$, or a single q-bosonic scalar field $\phi_q(\sigma)$, obtained from \textit{q}-deformed bosonization, can be used to explain the \textit{q}-deformed ghost sector. Eq. \eqref {01} provides the integration of a \textit{q}-deformed string field in the \textit{q}-deformed ghost sector after inserting $\exp_q(-3i\phi_q(\pi/2)/2)$ into the integral. While the \textit{q}-deformed exponential $\exp_q(-3i\phi_q(\pi/2)/2)$ is given as;
 \begin{equation}
\exp_q\Big(-3i\phi_q(\pi/2)/2\Big)=\sum_{m>0} \frac{\Big(-3i\phi_q(\pi/2)/2 \Big) ^m}{[m]_q!}
 \end{equation}

We can    represent the \textit{q}-deformed  string field $\Psi$ in terms of  scalar, vector and other fields as  
\begin{equation}\label{psi}
|\Psi \rangle_q= \int d^{26}p {[(\phi(p)_q + A_{ \mu}(p)_q \alpha^{\mu}_{-1}
 + \chi(p)_q \ :b_{-1} c_0:_q + \dots)]}|0;p \rangle_q ~,
\end{equation}
where $\phi(p)_q, A_{ \mu}(p)_q  $ are \textit{q}-deformed versions of scalar field $\phi{(p)}$ and vector field $\ A_{\mu}{(p)}$, respectively, with the \textit{q}-normal ordering on $ \ :b_{-1} c_0:_q$ given by  \eqref{BCM}. The 26-dimensional momentum integration is used to compute expectation values of physical observables, calculate scattering amplitudes, and determine the overall behavior of the string in spacetime. In the \textit{q}-deformed string theory context, this integration remains essential for capturing the quantum aspects and interactions of the \textit{q}-deformed oscillation modes.

 The algebraic structure for \textit{q}-deformed string field theory  action that is inspired by conformal field theory is given by
\begin{equation}\label{0p0}
    S_q= -\frac{1}{g^2}\Big[\frac{1}{2}\langle \Psi_q,{Q_q}\Psi_q\rangle + \frac{1}{3}\langle \Psi_q,\Psi_q \star \Psi_q \rangle\Big]~.
\end{equation}
where the \textit{g} is the open string coupling constant, the string field $\Psi$ is a state in the total matter plus ghost CFT, \textit{Q} is the BRST operator, $\star$ denotes a multiplication (star product). The \textit{q}-deformed BRST operator  $\textit{Q}_q$ satisfies the relation 
\begin{equation}
\langle \Psi_q  , {Q}_q \Psi_q  \rangle =(-1)^{\Psi_q} \langle {Q}_q \Psi_q  , \Psi_q \rangle 
\end{equation}
Here it may be noted that the action of \textit{q}-deformed BRST operator is similar to the action of the usual BRST operator in string field theory \cite{hh}. 
The \textit{q}-deformed star product $\star$ is equivalent to the multiplication of \textit{q}-deformed operators acting on half-string functionals $\Psi_q \star \Psi_q\Rightarrow \hat \Psi_q ~ \hat \Psi_q$.

The operators \(\hat{\Psi}_q\) are used to define the kinetic term of the \textit{q}-deformed string field theory action. When this kinetic term acts on a \textit{q}-deformed string state, it describes the evolution of the state with respect to the \textit{q}-deformed BRST operator. The construction of \(\hat{\Psi}_q\) ensures that this kinetic term captures the dynamics of \textit{q}-deformed string states that interact with the \textit{q}-deformed BRST operator.
In the ghost sector, the \textit{q}-star product has an extra insertion of  $\exp_q(3i \bar{\phi}/2)$ due to the ghost current anomaly. So, a pair of \textit{q}-deformed states are associated with their respective \textit{q}-deformed operators through \textit{q}-ghost functional $\Psi_q[\phi(\sigma)]$ with fixed ghost number. So, in Eq.\eqref{0p0}, the \textit{q}-deformed star product is defined as:
\begin{equation}\label{0o0}
\Psi_q \star \Psi_q = \hat{\Psi}_q \cdot M \cdot \hat{\Psi}_q
\end{equation}
it can further be written as;
\begin{equation}\label{0i0}
\Psi_q \star \Psi_q = \hat{\Psi}_q ~ e_q^{\left(\frac{3i}{2}\bar{\phi}\right)} ~ \hat{\Psi}_q
\end{equation}

 These two definitions are equivalent, which can be shown by expanding the $q$-deformed exponential term in equation \eqref{0o0} using its series expansion:
\begin{equation}
e_q^{\left(\frac{3i}{2}\bar{\phi}\right)} = \sum_{m>0}  \frac{\left(\frac{3i}{2}\bar{\phi}\right)^m}{[m]_q!}
\end{equation}

Upon substituting this expansion into equation \eqref{0i0} we have:
\begin{equation}
\Psi_q \star \Psi_q = \hat{\Psi}q \left(\sum_{m>0}  \frac{\left(\frac{3i}{2}\bar{\phi}\right)^m}{[m]_q!}\right) \hat{\Psi}_q
\end{equation}
where $q$-factorial $[m]_q!$ is defined as
\begin{eqnarray}
    [m]_q! &=& [1]_q\cdot [2]_q \cdot \cdot \cdot [m-1]_q\cdot[m]_q ~,\nonumber \\
    &=& \frac{1-q}{1-q} \cdot \frac{1-q^2}{1-q}\cdot \cdot \cdot \frac{1-q^{m-1}}{1-q} \cdot \frac{1-q^m}{1-q}~, \nonumber \\
    &=&(1)\cdot (1+q) \cdot \cdot \cdot (1+q+ \cdot \cdot \cdot +q^{m-2}) \cdot (1+q+ \cdot \cdot \cdot +q^{m-1})~.
\end{eqnarray}
The \textit{q}-deformed exponential series will expand to terms of different orders in $\bar{\phi}$ because it contains products of $\bar{\phi}$ raised to different powers. The \textit{q}-deformed binomial expansion can be used to represent these terms as a q-series. The important thing to take into account is that the q-series coefficients will be dependent on the \textit{q}-deformed factorials $[m]_q!$, which are also used to define the \textit{q}-deformed star product in equation \eqref{0p0}.

The fact that both formulations use identical q-series coefficients and \textit{q}-deformed factorials demonstrates that they are equivalent. Both definitions involve the operator $\hat{\Psi}_q$, and the \textit{q}-deformed midpoint factor $M$ affects the interaction between states in a way that is compatible with the \textit{q}-deformed exponential term. Due to the characteristics of the \textit{q}-deformed exponential and the q-series expansion, the mathematical analysis above shows that the two definitions of the \textit{q}-deformed star product are equivalent. This equivalence guarantees that the \textit{q}-deformed star product is consistent with our \textit{q}-deformed string field theory framework and that both definitions lead to the same mathematical structure.

Also, the inner product $\langle \Psi_q,\Psi_q \star \Psi_q \rangle$ can be defined in terms of a multilinear object that given three string fields yields a number;
\begin{equation}
   \langle  \Psi_q,\Psi_q \star \Psi_q \rangle \equiv \langle  \Psi_q,\Psi_q , \Psi_q \rangle
\end{equation}
The cyclicity property involving the inner product and star operation;
\begin{equation}
    \langle A, B \star C \rangle =  \langle A \star B, C \rangle 
\end{equation}
implies the cyclicity of the multilinear form. A small calculation immediately gives;
\begin{equation}
    \langle  \Psi_q,\Psi_q \star \Psi_q \rangle \equiv  \langle \Psi_q,\Psi_q , \Psi_q \rangle= (-)^{\Psi_q(\Psi_q +\Psi_q)}  \langle \Psi_q,\Psi_q , \Psi_q \rangle
\end{equation}
The \textit{q}-deformed string field $\Psi_q$ is again a functional of matter and ghost fields, the \textit{q}-deformed BRST operator ${Q}_q$ can be viewed as the \textit{q}-deformed kinetic operator.

\section{Properties of quantum deformed string field theory  }
 It may be noted that the \textit{q}-deformed Witten's string field theory satisfies the axioms of string field theory.  
 Before we try to investigate if the \textit{q}-deformed Witten's string field theory satisfies the axioms of a general string field theory, we will investigate the effect of \textit{q}-deformation on self-adjointness of the BRST operator. 
From the  inner product, we can observe that  
\begin{equation}
    \langle \Psi  , {Q}_q \Psi  \rangle_q=(-1)^{\Psi _q(1+\Psi _q)}\langle {Q}_q \Psi  , \Psi  \rangle_q =
    \langle {Q}_q \Psi  , \Psi  \rangle_q=-(-1)^{\Psi _q }  \langle \Psi  , {Q}_q \Psi \rangle_q ~.
\end{equation}
So ${Q}_q$ is self-adjoint, and the \textit{q}-string field $\Psi_q$ must be Grassmann odd.

The \textit{q}-deformed axioms of string field theory which are  satisfied by \textit{q}-deformed Witten's string field theory are:

\begin{itemize}
\item[1.] {  Nilpotency}:\ It is important to show that $Q^2_q \Psi_q = 0$ for every \textit{q}-deformed string field $\Psi_q$ in order to demonstrate the nilpotency of $Q_q$, which is previously confirmed in the earlier section provided in Eq. \eqref{qq2} to Eq. \eqref{pop4}

\item[2.] { Integration}:  $\int Q_q\Psi_q=0,  \  \ \forall \Psi_q \in \mathcal{A}$~. 

In split string operator language, we have;
\begin{equation}\label{arshid}
  \int Q_q\Psi_q= \mathrm{Tr} \Big[ M^{-1} \left( \hat{Q}_q \hat{\Psi}_q - (-1)^{G_{\Psi}} \hat{\Psi}_q 
  \hat{Q}_q \right) \Big]  
\end{equation}
The integral $ \int Q_q\Psi_q$ vanishes unless the ghost number of $\Psi_q$ equals two, and since we have;
\[ [ \hat{Q}_q,M]= [ \hat{\Psi}_q, M] =0\]
we see that Eq. \eqref{arshid} vanishes by cyclicity of the trace. For operators $\hat{\Psi}_q$ associated with well-behaved states this operation is valid, however. So a general BRST operator satisfies $ \int Q_q\Psi_q=0$ for well-behaved states $\Psi_q$.

\item[3.] {  Derivation}:   ${Q}_q (\Psi _q \star \Phi_q)= ({Q}_q \Psi _q) \star \Phi _q + (-1)^{G \Psi _q} \Psi _q \star ({Q}_q \Phi_q)$~.
which defines the derivative property of ${Q}_q$. Here we will use the operator formslism, where $Q_q(\Psi_q)=Q_q\hat\Psi_q -(-)^{G_{\Psi_q}}\hat \Psi_q Q_q$.
By taking the expression of \textit{q}-deformed star product into consideration, i.e.
\begin{equation}\label{qw}
    \Psi \star \Phi = \hat \Psi M \hat \Phi~,
\end{equation}
we have
\begin{equation}\label{mid1}
   {Q}_q (\Psi _q \star \Phi _q) \equiv  {Q}_q \hat\Psi _q M \hat\Phi _q -(-1)^{G_{\Psi_q + \Phi_q} }\hat\Psi _q M \hat\Phi _q {Q}_q~.
\end{equation}
On the other hand,
\begin{eqnarray}\label{mid2}
     {Q}_q (\Psi _q) \star \Phi _q + (-1)^{\Psi _q} \Psi _q \star {Q}_q( \Phi _q)&=&( Q_q \hat \Psi _q -(-1)^{G_{\Psi _q}} \hat \Psi _q  Q_q)M \hat \Phi _q \nonumber \\
     &+& (-1)^{G_{\Psi _q}} \hat \Psi_q M( Q_q \hat\Phi _q -(-1)^{G_{\Phi _q}} \hat \Phi _q  Q_q).
\end{eqnarray}
Using the fact that $ Q_q$ commutes with $M$,  we can see that Eqs. \eqref{mid1} and \eqref{mid2} are indeed equal to each other. 
\item[4.] {Cyclicity}:  $\int \Psi _q \star \Phi_q= (-1)^{G_{\Psi_q}G_{\Phi_q}} \int \Phi_q \star \Psi_q, \  \ \forall  \{\Psi_q, \Psi_q  \}\in\mathcal{A}   $~.
 
The integral of star product of two fields is related to their inner product as 
\begin{equation}\label{inner}
    \int \Psi_q \star \Phi_q= \langle  \Psi_q, \Phi_q  \rangle~.
\end{equation}
 It should be noted that the inner product defines the symmetric condition which is given by
 \begin{equation}\label{10}
     \langle  \Psi_q, \Phi_q  \rangle= (-)^{G_{\Psi_q} G_{\Phi_q}} \langle  \Phi_q, \Psi_q  \rangle~.
 \end{equation}
Comparing Eqs. \eqref{inner} and \eqref{10}, and on rearranging we will obtain;
\begin{equation}
    \int  \Psi_q \star \Phi_q= (-1)^{G_{\Psi_q}G_{\Phi_q}} \int \Phi_q \star \Psi_q~.
\end{equation}
\item[5.]  {Associativity}: $
(\Psi_q \star \Phi_q) \star \Xi_q= \Psi_q \star (\Phi_q \star \Xi_q).$

Here we can  use the expression 
$ (\Psi_q \star \Phi_q)= \hat\Psi_q M \hat\Phi_q$ to demonstrate that the  \textit{q}-deformed star product is associative. We first observe that 
\begin{equation}
   ( \Psi_q \star \Phi_q) \star \Xi_q= (\hat \Psi_q M \hat\Phi_q) \star \hat\Xi_q=\hat\Psi_q M \hat\Phi_q M \hat \Xi_q
\end{equation}

Using the  same procedure, we will also have; 
\begin{equation}
\Psi_q \star( \Phi_q\star \Xi_q)=  \hat \Psi_q M \hat \Phi_q M\hat \Xi_q.
\end{equation}
So, the \textit{q}-deformed star product is associative. 
\end{itemize}
Hence, the \textit{q}-deformed string field theory satisfied the axioms of string field theory.

\section{Quantum deformed string  vertices  }
For many computations, it is useful to think of the \textit{q}-deformed string as being ``split” into a left half and a right half. Formally, the \textit{q}-deformed string field can be expressed as a functional $\Psi_q[L, R]$, where $L, R$ describe the left and right parts of the \textit{q}-deformed string. This is a very appealing idea since it leads to
a simple picture of the \textit{q}-deformed star product in terms of matrix multiplication. Nonetheless, some of the structure of the \textit{q}-deformed star product encoded in the \textit{q}-deformed three-string vertex
will be the easiest way to understand using the half-string formalism, and many formulae related to the \textit{q}-deformed 
3-string vertex can easily expressed in terms of the linear map from full-string modes
to half-string modes.

To perform explicit computations in open string field theory,  we need to consider interactions of the fields (other than the tachyon).  Such calculations can be done in Fock space representations of the string vertices.  In this section, we will explicitly construct  \textit{q}-deformation of such string vertices. 
The vertex ($\langle V_{N}| $, $N = 1, 2, 3$) is a solution of overlap conditions \cite{hh}
\begin{eqnarray}
   \langle V_{N}|[X^{(i)} (\sigma) - X^{(i-1)} (\pi - \sigma)] =0~, \nonumber \\ 
    \langle V_{N}|[c^{(i)} (\sigma) + c^{(i-1)} (\pi - \sigma)] =0~, \nonumber \\ 
     \langle V_{N}|[b^{(i)} (\sigma) - b^{(i-1)} (\pi - \sigma)] =0~.
\end{eqnarray}
Here, the solutions to the overlap conditions represent the interactions between respective strings. Now  $i= 1,2,...,N$ denotes the number of strings and $X(\sigma) = X(\sigma ,0)$, $c(\sigma) = c(\sigma ,0), b(\sigma) = b(\sigma, 0)$,   $X(\sigma ,\tau), c(\sigma ,\tau) $ and $b(\sigma, \tau)$, which are  given by 
\begin{eqnarray}
    X^\mu(\sigma,\tau) &=& x^\mu + 2p^\mu \tau + i \sqrt{2 \alpha^\prime} \sum_{n=1}^{\infty} \frac{1}{n} {\alpha_n ^\mu}cos(n\sigma) e^{-i n \tau} ~,\nonumber \\
    c_{\pm}(\sigma,\tau ) &=&\sum_{m>0}  c_m e^{-im(\tau \pm \sigma)}~, \nonumber \\ 
   b_{\pm \pm}(\sigma,\tau ) &=&\sum_{m>0}  b_m e^{-im(\tau \pm \sigma)}~.
\end{eqnarray}
Two string-vertex  overlapping conditions can be used to obtain the following expression for    the overall formula  \cite{hh}
\begin{eqnarray}\label{..}
    \langle V_2| = \int d^{26}p(
    \langle 0;p| \otimes \langle 0;-p|) (c_0^{(1)} + c_0^{(2)}) \nonumber \\
    \times \exp \Big( - \prod_{m>0}  (-1)^m \big[ \frac{1}{m}\alpha_m^{(1)} \alpha_m^{(2)} + c_m^{(1)}b_m^{(2)} + c_m^{(2)}b_m^{(1)} \big] \Big) ~.
\end{eqnarray}
The \textit{q}-deformation of  Eq. \eqref{..} can be expressed as 
\begin{eqnarray}
   _q \langle V_2| = \int d^{26}p(
    _q\langle 0;p| \otimes \  _q\langle 0;-p|) (c_0^{(1)} + c_0^{(2)}) \nonumber \\
    \times \exp_q \Big( - \prod_{m>0} (-1)^{[m]} \Big[ \frac{1}{[m]}\alpha_{[m]}^{(1)} \alpha_{[m]}^{(2)} + c^{(1)}_{[m]}b^{(2)}_{[m]} + c^{(2)}_{[m]}b^{(1)}_{[m]} \Big] \Big) ~.
\end{eqnarray}
Here $ _q\langle 0;-p| \equiv \ _q\langle 0 |c_{-1} \otimes \ _{M_q}\langle -p|$ with $M_q$  the \textit{q}-deformed matter component of vacuum, also the \textit{q}-exponential function  is given by  
\begin{eqnarray}
\exp_q \Big( - \prod_{m>0}  (-1)^{[m]} \Big[ \frac{1}{[m]}\alpha_{[m]}^{(1)} \alpha_{[m]}^{(2)} + c^{(1)}_{[m]}b^{(2)}_{[m]} + c^{(2)}_{[m]}b^{(1)}_{[m]} \Big] \Big) = \nonumber \\
 \sum_{m=0}^{\infty} {\Big( - \prod_{m>0}  (-1)^{[m]} \Big[ \frac{1}{[m]}\alpha_{[m]}^{(1)} \alpha_{[m]}^{(2)} + c^{(1)}_{[m]}b^{(2)}_{[m]} + c^{(2)}_{[m]}b^{(1)}_{[m]} \Big] \Big)^m }{([m]_q!)^{-1}}~. \end{eqnarray}
 where $c^{(1)}_{[m]}b^{(2)}_{[m]}$ and $c^{(2)}_{[m]}b^{(1)}_{[m]}$ are given as
 \begin{eqnarray}
    c^{(1)}_{[m]}b^{(2)}_{[m]} = -\frac{1}{q^{N_m}} b^{(2)}_{[m]}c^{(1)}_{[m]}, &&
    c^{(2)}_{[m]}b^{(1)}_{[m]}= -\frac{1}{q^{N_m}}b^{(1)}_{[m]}c^{(2)}_{[m]}~.
 \end{eqnarray}

Now, we can consider the case of three interacting strings.    The \textit{q}-deformed three-string vertex, which is associated with the three-string overlap, can be computed in a very similar way to the \textit{q}-deformed two-string vertex. The general formula for the three-string vertex has already been obtained in \cite{hh}, here we are generalizing it to the \textit{q}-deformed version of those interactions. Thus, repeating the calculations used to obtain such string vertices, for \textit{q}-deformed string vertices, we have 
\begin{eqnarray}
    _q\langle V_3| = \kappa {\prod_{i=1}^3} \Big(\int d^{26}p^{(i)}|0;p^{(i)}\rangle_q \Big) \delta_q(p^{(1)} + p^{(2)} + p^{(3)}) {c_{0}^{(1)}}  {c_{0}^{(2)}} {c_{0}^{(3)}} \nonumber \\
    \times \exp_q \Bigg( -\frac{1}{2} {\sum_{r,s=1}^3} \prod_{m,n} \Big[\frac{1}{\sqrt{[m][n]}}\alpha_{[m]}^{(r)} {V_{[m][n]}^{rs}}\alpha_{[n]}^{(s)}  + 2 \alpha_{[m]}^{(r)} \frac{V_{[m][0]}^{rs}}{\sqrt{[m]}}  p^{(s)} \nonumber \\ + p^{(r)} {V_{[0][0]}^{rs}} p_{[n]}^{(s)}+ c_{[m]}^{(r)} X_{[m][n]}^{rs} b_{[n]}^{(s)} \Big] \Bigg)~.
\end{eqnarray}
where $\kappa =\mathcal{N} =K^3 ={3^{9/2}}/2^6$,
and \textit{q}-exponential function is given by 

\begin{eqnarray}
 &\exp_q \Bigg( -\frac{1}{2} {\sum_{r,s=1}^3} \prod_{m,n} \Big[\frac{1}{\sqrt{[m][n]}}\alpha_{[m]}^{(r)} {V_{[m][n]}^{rs}}\alpha_{[n]}^{(s)}  + 2 \alpha_{[m]}^{(r)} \frac{V_{[m][0]}^{rs}}{\sqrt{[m]}}  p^{(s)}+ p^{(r)} {V_{[0][0]}^{rs}} p_{[n]}^{(s)}+ c_{[m]}^{(r)} X_{[m][n]}^{rs} b_{[n]}^{(s)} \Big] \Bigg)  \nonumber \\ &= 
 \sum_{m=0}^{\infty}\Bigg[{ \Bigg( -\frac{1}{2} {\sum_{r,s=1}^3} \prod_{m,n} \Big[\frac{1}{\sqrt{[m][n]}}\alpha_{[m]}^{(r)} {V_{[m][n]}^{rs}}\alpha_{[n]}^{(s)}  + 2 \alpha_{[m]}^{(r)} \frac{V_{[m][0]}^{rs}}{\sqrt{[m]}}  p^{(s)}+ p^{(r)} V_{[0][0]}^{rs}} p_{[n]}^{(s)} \nonumber \\  &+ c_{[m]}^{(r)} X_{[m][n]}^{rs} b_{[n]}^{(s)} \Big] \Bigg)^{m}{([m]_q!)^{-1}}\Bigg]~.
\end{eqnarray}
 Here   ${V_{[m][n]}^{rs}}$ are the Neumann coefficients and $ X_{[m][n]}^{rs}$ are constants. The delta functions occur due to the overlap of the three strings, and they also contain the ghost sector. In the limit, $q \rightarrow 1$:  the \textit{q}-deformed string vertices reduce to the undeformed string vertices. \\

 \section{Quantum-deformed Witten Vertex}

Witten's vertex, is a fundamental component in the framework of open string field theory, particularly in the context of string theory. It represents the three-string interaction term in the open string theory action and plays a crucial role in understanding the dynamics of open strings.
 Witten's cubic vertex operator captures the non-perturbative interactions between open strings in string theory and is an essential ingredient for constructing the full open string field theory action.
Witten's cubic open string field theory vertex, denoted as \(V_3^W(A,B,C)\), describes the interaction between three open string states. In mathematical terms, it can be represented as:

\begin{equation}
V_3^W(A, B, C) = \langle : f_{(3,1)}^W \circ A(0) \, f_{(3,2)}^W \circ B(0) \, f_{(3,3)}^W \circ C(0) : \rangle_{UHP}
\end{equation}

\(A(0)\), \(B(0)\), and \(C(0)\) are the vertex operators associated with the three open string states involved in the interaction.  \(f_{(3,1)}^W\), \(f_{(3,2)}^W\), and \(f_{(3,3)}^W\) are conformal maps that transform the locations of these vertex operators in the complex plane. These maps are chosen such that they map the half-disk to a \(120^{\circ}\) wedge of the unit disk \cite{1a}.  These conformal maps are designed to simplify the calculation of the three-string interaction amplitude. The conformal transforms are defined by the conformal mapping as
\begin{equation}
    f_a(z_a) =\alpha^{2-a} f(z_a)_{a=1,2,3}~,
\end{equation}
where $\alpha = e^{2\pi i/3}$ is defined to be the puncture. The punctures of the first, second, and third state are fixed to lie at either  $a=1$, $a=0$ or $a=\infty$.

Now the \textit{q}-deformation of Witten's cubic vertex involves introducing the deformation parameter \(q\) into the original conformal maps for the vertex. The \textit{q}-deformed versions of the conformal maps \(f_{(3,1)}^W(\xi)_q\), \(f_{(3,2)}^W(\xi)_q\), and \(f_{(3,3)}^W(\xi)_q\) are as:

\begin{subequations}
    \begin{eqnarray}
        \Big(f_{(3,1)}^W(\xi)\Big)_q = \frac{1 - e_q^{2\pi i/3}\left(\frac{1 + i\xi}{1 - i\xi}\right)^{2/3}}{\left(\frac{1 + i\xi}{1 - i\xi}\right)^{2/3} - e_q^{2\pi i/3}}\\
        \Big(f_{(3,2)}^W(\xi)\Big)_q = \frac{1 - \left(\frac{1 + i\xi}{1 - i\xi}\right)^{2/3}}{e_q^{-2\pi i/3}\left(\frac{1 + i\xi}{1 - i\xi}\right)^{2/3} - e_q^{2\pi i/3}}\\
         \Big(f_{(3,3)}^W(\xi)\Big)_q = \frac{e_q^{-2\pi i/3} - e_q^{2\pi i/3}\left(\frac{1 + i\xi}{1 - i\xi}\right)^{2/3}}{\left(\frac{1 + i\xi}{1 - i\xi}\right)^{2/3} - 1}
    \end{eqnarray}
\end{subequations}

In these expressions, the parameter \(q\) introduces a deformation in the conformal maps compared to the original conformal maps. These \textit{q}-deformed maps are used in the context of \textit{q}-deformed string theory and play a role in calculations related to the interactions of \textit{q}-deformed open string states.
Here, again, \textit{q}-exponential functions will be given 
by; 
\[\exp_q{(\pm2\pi i/3)} =\sum_{m>0} \frac{(\pm 2\pi i/3)^m}{[m]_q!}\]
This specification of the \textit{q}-deformed conformal  functions gives the definition of \textit{q}-deformed cubic vertex, with puncturtes $e^{2\pi i/3},1$ and $e^{-2\pi i/3}$. We need to contrast the two calculations to determine whether the \textit{q}-deformed Witten vertex and the \textit{q}-deformed cubic vertex computed using the operator formulation are equivalent.
It is now important to demonstrate that these two expressions are equivalent for appropriate values of \(A(0)_q\), \(B(0)_q\), and \(C(0)_q\) to verify their equivalence. We must specifically show that the exponential term in the \textit{q}-deformed cubic vertex fits the correlator in the upper half-plane when evaluated with the \textit{q}-deformed conformal maps. 

\section{Conclusion}
We first reviewed the construction of     \textit{q}-deformation of string theory in a covariant gauge. Thus, we reviewed the construction of  \textit{q}-deformation of string states, normal ordering, Virasoro algebra, and the BRST operator. This was then used to construct a   \textit{q}-deformation of Witten's open string field theory. It was explicitly demonstrated that the \textit{q}-deformed Witten's cubic string field theory satisfied all the axioms of a general string field theory. This \textit{q}-deformed cubic string field theory was finally used to obtain a deformation of string vertices.   The work of this paper can be used to analyze aspects of \textit{q}-deformation of Vasiliev theory.
 String theory expanded around Minkowski's background takes the form of a massive higher spin theory, which contains towers of fields lying on Regge trajectories.  It has been proposed that the   Minkowski vacuum spontaneously breaks a  higher spin gauge symmetry of this theory \cite{hs}. This symmetry is expected to become manifest in the tensionless limit. In such a limit, all the fields are expected to organize themselves in multiple of the symmetry. As the higher spin gauge theories have been analyzed on an AdS background   \cite{hs12}, it has been proposed that the tensionless limit of string theory can be studied on  AdS spacetime using this higher spin gauge theories \cite{hd14}. Such a tensionless string field theory on  AdS spacetime has also been studied using  Vasiliev theory \cite{hs16}. It would be interesting to study such a tensionless limit of \textit{q}-deformed string field theory on AdS spacetime. It would also be interesting to analyze if this can be related to   \textit{q}-deformation of Vasiliev theory.

 %While deriving the second quantized field theory we have two ways to approach. The first method is the ``bottom-up" approach  which tries to guess the action from first quantized theory. And the success of this approach is we are able to write down the BRST action which is the first covariant field theory of strings.And the second method `` top-down" approach which is strictly based on geometry and group theory. Which follows close analogy with Yang-Mills theory and General relativity. It would be intresting to study ``top-down" approach under the \textit{q}-deformation of  Yang-Mills theory. Since tachyons having $(mass)^2$ less than some critical value can generate a deformation in the original CFT by an exactly marginal operator that describes a rolling tachyon solution\cite{.0}. It would be interesting to study  the \textit{q}-deformed theory for rolling tachyon solutions which will not require any additional renormalization to renormalize the original CFT.  The quantum open-closed homotopy algebra is an  algebraic structure associated with quantum open-closed string field theory \cite{0.0},  it would be interesting to study the modifications to this   algebra from  \textit{q}-deformations.

The tachyonic instability can be addressed using condensation in string field theory  \cite{a8}.
Due to tachyon condensation, such a system with tachyons can become stable. It is possible to obtain a  configuration with zero energy density for a system containing a tachyonic mode along with a  brane and an anti-brane      \cite{8}. Here,  the negative energy density associated with the tachyonic potential is exactly canceled by the tension of the brane and the anti-brane. As the expansion of the string field gets modified due to  \textit{q}-deformation, it is expected that this will modify the negative energy density associated with the tachyonic potential. However, it is possible that this modified negative energy density associated with the tachyonic potential will still be canceled by the modifications of the brane and the anti-brane system by \textit{q}-deformation. 
It would be interesting to investigate such a  \textit{q}-deformation of tachyonic condensation in string theory.   The implications of string field theory on cosmology have been investigated using a novel method to deal with nonlocal Friedmann equations   \cite{Arefeva:2005mom, Joukovskaya:2007nq}. It was observed that such nonlocal Friedmann equations obtained from string field theory produce non-singular bouncing cosmological solutions. The generalization of this analysis to inflation has also been done using this novel nonlocal formalism \cite{inflation}. It is expected that the \textit{q}-deformation will modify the nonlocal Friedmann equations, and this will modify the cosmological solutions obtained from these equations. These modifications will also modify the inflationary cosmology obtained from string field theory. It will be interesting to first analyze the effect of \textit{q}-deformation on such nonlocal Friedmann equations, and then analyze the effect on inflation.\\ 
The exploration of \textit{q}-deformation in string field theory stems from a confluence of theoretical interests and potential implications for fundamental physics. This deformation introduces a novel mathematical framework that allows us to probe non-standard deformations of conventional string theory, thereby opening avenues for investigating a broader spectrum of physical phenomena. Below are some key motivations for delving into \textit{q}-deformed string field theory.

 Quantum groups and quantum algebras provide the mathematical foundation for the idea of \textit{q}-deformation. We introduce non-commutative structures by applying the fundamentals of quantum mechanics to algebras, opening up fascinating new perspectives on the behavior of strings both at the classical and quantum levels. The algebraic relations regulating the fundamental degrees of freedom of the string experience non-trivial modification due to the deformation of string oscillator modes. This makes it possible to investigate how these deformations affect the string's vibrational modes, possibly leading to new physical hypotheses.
   We may investigate the effects of quantum symmetries in the context of string dynamics owing to the \textit{q}-deformation of the Virasoro algebra, a key component of string theory. It is clearer to understand the interaction between symmetries and quantum effects when one is aware of how these deformations impact the algebraic structure.
   Conventional string theory deformations can provide a glimpse into possible deviations from fundamental quantum physics at the Planck scale. We may learn more about the nature of spacetime at these incredibly small scales by examining the effects of \textit{q}-deformation.
   The field of non-commutative geometry, which has attracted considerable interest from both mathematicians and theoretical physicists, is a perfect fit for the \textit{q}-deformation framework. This linkage might open up unanticipated dualities and serve as a link between string theory and other non-commutative theories.
  There may be consequences for the broader pursuit of a quantum theory of gravity if the implications of \textit{q}-deformation in string field theory are understood. It provides a way to investigate several quantum-level theories of gravity, maybe illuminating the fundamental properties of spacetime.

In conclusion, the study of \textit{q}-deformation in string field theory offers a fertile context for examining unusual deformations of fundamental physics. It provides access to a wide range of potential discoveries, from fresh mathematical formulations to new understandings of the fundamental properties of spacetime at the quantum level.

\section*{Acknowledgement} We would like to thank Meer Ashwinkumar for the useful discussion. On behalf of all authors, the corresponding author states that there is no conflict of interest.


\begin{thebibliography}{200}
 
\bibitem{q101} I V Volovich, Class. Quantum Grav. 4 L83 (1987).


\bibitem{q7}M.~Frasca, A.~Ghoshal and N.~Okada, Phys. Rev. D  {104} (2021)  096010.
\bibitem{q8} P.~H.~Frampton, J. Phys. A {53} (2020)   191001.

\bibitem{q1} C.~S.~Chu and P.~M.~Ho, Nucl. Phys. B  {636} (2002) 141. 

\bibitem{q2} Martin Schnabl, JHEP11(2000), 031. 
\bibitem{q3} D.~Roychowdhury, Phys. Rev. D  {95} (2017)  086009. 
\bibitem{q4}S.~Hellerman, D.~Orlando and S.~Reffert, JHEP  {01} (2012) 148.
\bibitem{q5a} S.~Hirano, T.~Nakajima and M.~Shigemori, JHEP  {04} (2021) 270.
\bibitem{kappa}F. Delduc, M. Magro, and B. Vicedo,
Phys. Rev. Lett. 112 (2014) 051601.
\bibitem{first}M.~Chaichian, J.~F.~Gomes and P.~Kulish, Phys. Lett. B {311} (1993) 93.
 

\bibitem{first14} 
M.~Chaichian and P.~Presnajder, Nucl. Phys. B {482} (1996) 466.

\bibitem{n0d} Jurg Frohlich and Thomas Kerler, \textit{Quantum groups, quantum categories and quantum field theory}  \textbf{Lect. Notes Math.} (1993) 1542.
\bibitem{n1d}Eremin, V.V., Meldianov, A.A, Theor Math Phys 147, (2006), 710.
\bibitem{n2d} P. P. Raychev, R. P. Roussev, Nicola Lo Iudice and P. A. Terziev,  J. Phys. G Nucl. Part. Phys.  {24} (1998) 1931.
\bibitem{functionss} L.~B.~Drissi, H.~Jehjouh and E.~H.~Saidi, Nucl. Phys. B {801} (2008) 316.
%\bibitem{l1}E.~Vernier, J.~L.~Jacobsen and J.~Salas, J. Phys. A  {49} (2016) 174004. 
  
\bibitem{l3}S.~Dasmahapatra, R.~Kedem, T.~R.~Klassen, B.~M.~McCoy and E.~Melzer, Int. J. Mod. Phys. B  {7} (1993) 3617. 
 
\bibitem{yangb}
V.~Belavin and D.~Gepner,
JHEP  {02},  (2019)  033.
 
\bibitem{Z2}E. K. Sklyanin, Funct. Anal. Appl. 16 (1982) 27.
 
\bibitem{Z4}M. Jimbo, Lett. Math. Phys. 10 ( 1985 ) 63.

\bibitem{dualties}T.~Matsumoto and K.~Yoshida,
 JHEP  {03} (2015) 137.
 \bibitem{mtheory}E.~Malek, Y.~Sakatani and D.~C.~Thompson, JHEP  {01} (2021) 020. 
 \bibitem{Baxter}S.~Hronek and L.~Wulff, JHEP  {10} (2020) 065. 
 
 \bibitem{Yang}A.~\c{C}atal-\"Ozer and S.~Tunal, Class. Quant. Grav.  {37} (2020)   075003. 

\bibitem{Z5}U Alvarez-Gaum, C. Gomez and G. Sierra, Phys. Lett. B 220 (1989) 142.  
\bibitem{diracc}M.~Lotfizadeh, J. Math. Phys. {61} (2020)   063502.
\bibitem{multi} K.~D.~Sviratcheva, A.~I.~Georgieva and J.~P.~Draayer, J. Phys. A  {36} (2003) 7579.
\bibitem{Z6}T. Curtright and C. Zachos, Phys. Lett. B 243 (1990) 237.
\bibitem{Z7}M. Chaichian, P.P. Kulish and J. Lukierski, Phys. Lett. B 237 (1990 ) 401.
\bibitem{Z8}M. Chaichian, D. Ellinas and Z. Popowicz, Phys. Left. B 248 (1990) 95.


\bibitem{dual1} I.~Ridkokasha,
Mod. Phys. Lett. A  {36} (2021)   2150031.

\bibitem{dual2}  M.~Chaichian, J.~F.~Gomes and R.~Gonzalez Felipe,
 Phys. Lett. B  {341} (1994)  147.
 
\bibitem{kl} J. M. Maldacena,   Int. J. Theor. Phys.
38 (1999) 1113.  
\bibitem{lk} N. Beisert et al.,  Lett. Math. Phys. {99} (2012) 3.

\bibitem{inte0}J.~Kluson, B.~H.~Lee, K.~L.~Panigrahi and C.~Park,
 JHEP  {08} (2008)  032.
\bibitem{inte01} J.~A.~Minahan, A.~Tirziu and A.~A.~Tseytlin, JHEP  {08} (2006)  049. 
\bibitem{inte1}N.~Beisert, C.~Ahn, L.~F.~Alday, Z.~Bajnok, J.~M.~Drummond, L.~Freyhult, N.~Gromov, R.~A.~Janik, V.~Kazakov and T.~Klose, \textit{et al.}
 Lett. Math. Phys.  {99} (2012)  3.
 \bibitem{inte2} N. Beisert and P. Koroteev, J. Phys. A 41 (2008) 255204.
\bibitem{inte4}N. Beisert, J. Phys. A 44 (2011) 265202.
\bibitem{inte5}B. Hoare and A. A. Tseytlin, Nucl. Phys. B851 (2011) 161.
\bibitem{inte6}B.~Hoare, T.~J.~Hollowood and J.~L.~Miramontes,
 JHEP  {11} (2011) 048.
\bibitem{smatrix}B.~Hoare, T.~J.~Hollowood and J.~L.~Miramontes,
 JHEP  {03} (2012), 015.
 \bibitem{smatrix1}G. Arutyunov, S. Frolov and M. Staudacher, JHEP 0410 (2004) 016.
 \bibitem{intea1}G. Arutyunov and S. Frolov, JHEP 0712 (2007) 024.
 \bibitem{intea2} G. Arutyunov and S. Frolov, JHEP 0911 (2009) 019.

\bibitem{xz12}F.~Delduc, M.~Magro and B.~Vicedo,
 JHEP  {10} (2014)  132.
\bibitem{XZ}  T. Kameyama and K. Yoshida, J. Phys. A {48} (2015) 075401.


 
\bibitem{B1} D. E. Berenstein, R. Corrado, W. Fischler and J. M. Maldacena,  Phys. Rev. D {59} (1999) 105023.
\bibitem{C1}  N. Drukker, D. J. Gross and H. Ooguri, JHEP {1106} (2011) 131.

\bibitem{T4} F.~K.~Seibold, S.~J.~van Tongeren and Y.~Zimmermann, JHEP  {09} (2021) 110. 
 \bibitem{wilson} N.~Bai, H.~H.~Chen and J.~B.~Wu,
 Chin. Phys. C  {39} (2015) 103102.
\bibitem{sigma}T.~Kameyama and K.~Yoshida, JHEP  {08} (2014), 110.
\bibitem{Erler:2019loq}
T.~Erler, Phys. Rept.  {851} (2020) 1.

\bibitem{shell1}V.~Forini, G.~Grignani and G.~Nardelli, JHEP {04} (2006), 053. 
\bibitem{shell2} B.~Urosevic, Phys. Rev. D {50} (1994) 4075.
  \bibitem{back1}A.~Sen and B.~Zwiebach, Nucl. Phys. B  {414} (1994) 649. 
  \bibitem{back2}A.~Sen, JHEP {02} (2018) 155. 


\bibitem{sft01}A.~Sen, JHEP {12} (2015) 075.
\bibitem{sft02} A.~Sen, JHEP {11} (2016) 050. 
  \bibitem{prob6}A.~Sen, JHEP {12} (2016) 115. 
  \bibitem{prob7} C.~De Lacroix, H.~Erbin and A.~Sen, JHEP  {05} (2019) 139. 

\bibitem{witten}  David J. Gross and Washington Taylor JHEP {08} (2001) 010.

\bibitem{lightcone} T.~Lee,
 Phys. Lett. B  {768} (2017), 248.
 \bibitem{defo1}O.~K.~Kwon, Nucl. Phys. B {804} (2008) 1.
 
 \bibitem{defo2} S.~H.~Lai, J.~C.~Lee, T.~Lee and Y.~Yang, Phys. Lett. B {776} (2018) 150.
 
\bibitem{mu}C.~Maccaferri and J.~Vo\v{s}mera,
JHEP {09} (2021) 047. 

\bibitem{sine}I.~Kishimoto, T.~Kitade and T.~Takahashi, PTEP {2018} (2018)  123B04.

\bibitem{margin}F.~Katsumata, T.~Takahashi and S.~Zeze, JHEP {11} (2004) 050.

\bibitem{mass}C.~Maccaferri and J.~Vo\v{s}mera,
JHEP {09} (2021) 049.
 
 \bibitem{inte} K. D. Sviratcheva, C. Bahri, A. I. Georgieva, and J. P. Draayer
Phys. Rev. Lett. 93, (2004) 152501.

\bibitem{CC} Bianchi, Massimo and Firrotta, Maurizio, Nucl. Phys. B 952 (2020) 114943.
\bibitem{Biedenharn} Haruo Ui and N. Aizawa, Modern Phys. Letters A  {5}  (1990) 240.
\bibitem{vira string} Schellekens, A. N, Fortsch.Phys.  {44} (1996) 650.
\bibitem{V1} T. L. Curtright and C. K. Zachos, Phys. Lett. B  {243} (1990) 237.
\bibitem{V2} M. Chaichian, P. P. Kulish and J. Lukierski, Phys. Lett. B  {237} (1990) 401.
\bibitem{V3} M. Chaichian, A. P. Isaev, . Lukierski, Z. Popowicz and P. Presnajder, Phys. Lett. B  {262} (1991) 32. 
%\bibitem{V5} N. Aizawa and H. Sato, Phys. Lett. B  {256} (1991) 185.
\bibitem{V4} Aizawa, N., and Sato, H. Physics Letters B, 256(2), 1991, 187.

 \bibitem{NGT}Mebarki, N. and Boudine, A. and Aissaoui, H. and Benslama, A. and Bouchareb, A. and Maasmi, A., J. Geom. and Phys.  {30} (1999) 110.  
\bibitem{VA} Harutada SATO, Progress of Theoretical Physics,  {89} (1993) 540.
\bibitem{i1} M. Kaku and K. Kikkawa, Phys. Rev. D {10} (1974) 1823.
\bibitem{i2} E. Cremmer and J. L. Gervais, Nucl. Phys. B {76} (1974) 1100.
\bibitem{i3}  E. Cremmer and J. L. Gervais, Nucl. Phys. B {90} (1975) 410.
\bibitem{sft0} Taylor, Washington and Zwiebach, Barton, "D-branes, tachyons, and string field theory", \textit{Theoretical Advanced Study Institute in Elementary Particle Physics (TASI 2001): Strings, Branes and EXTRA Dimensions}",
\textbf{eprint = hep-th/0311017},  (2003). 
  
\bibitem{hh} Arefeva, Irina Ya. and Gorbachev, R. V. and Medvedev, P. B. and Rychkov, D. V., Three lectures on (super)string field theories,
\textit{3rd Summer School in Modern Mathematical Physics}, {2004}, 230.

 
\bibitem{GH} Malik, R. P.,
    Mod. Phys. Lett. A, \textbf{11}, (1996), 2871-2881, \textit{e-Print: hep-th/9503025 [hep-th]}.
\bibitem{1a}T.~Erler and H.~Matsunaga, JHEP  {11} (2021) 208.
\bibitem{hs}Gross, David J,  Phys. Rev. Lett. 60 (1988) 1229.
\bibitem{hs12} M. A. Vasiliev,   Phys. Lett. B 243  (1990) 378.
\bibitem{hd14} B. Sundborg,  
Nucl. Phys. Proc. Suppl. 102 (2001) 113.
\bibitem{hs16}J.~Raeymaekers, JHEP {07} (2019) 019.
\bibitem{a8} B.~Zwiebach, JHEP  {09}  (2000) 028.
\bibitem{8}A. Sen, JHEP  {08} (1998)  012.
\bibitem{Arefeva:2005mom}
I.~Aref'eva and L.~Joukovskaya, JHEP  {10} (2005)  087.
\bibitem{Joukovskaya:2007nq}
L.~Joukovskaya, Phys.\ Rev.\ D {76}  (2007) 105007.
\bibitem{inflation}H.~Sheikhahmadi, M.~Faizal, A.~Aghamohammadi, S.~Soroushfar and S.~Bahamonde,
Nucl. Phys. B {961} (2020) 115252.
\bibitem{ar2} H. Hiro-oka, O. Matsui, T. Naito and S. Saito, Tokyo Metropolitan Univ. Preprint TMUP-HEL- 9004 (1990).
\end{thebibliography}
\end{document}